\documentclass[12pt,reqno]{amsart}
\usepackage{lineno} 
\usepackage[utf8]{inputenc}
\usepackage{lmodern}
\fontfamily{lmss}\selectfont
 \usepackage{cancel} 

\usepackage{latexsym,amsmath,amscd,amssymb,graphics,cite}
\usepackage{enumerate}
\usepackage{todonotes}
\usepackage{comment}
\usepackage[margin=3cm]{geometry}

\usepackage{graphicx}
\usepackage{framed}
\usepackage[colorlinks]{hyperref}
\usepackage{url}
\usepackage{cite}

\usepackage[all]{xy}

\newcommand{\bfi}{\bfseries\itshape}

\makeatletter

\@addtoreset{figure}{section}
\def\thefigure{\thesection.\@arabic\c@figure}
\def\fps@figure{h, t}
\@addtoreset{table}{bsection}
\def\thetable{\thesection.\@arabic\c@table}
\def\fps@table{h, t}
\@addtoreset{equation}{section}

\makeatother

\newtheorem{theorem}{Theorem}

\newtheorem{remark}[theorem]{Remark}

\numberwithin{theorem}{section}

\def\be{\begin{equation}}
\def\ee{\end{equation}}
\def\bea{\begin{eqnarray}}
\def\eea{\end{eqnarray}}
\def\ba{\begin{array}}
\def\ea{\end{array}}

\def\bOm{\boldsymbol{\Omega}}

\def\bM{{\bf M}}

\def\m{\mu}

\def\d{\sf d}

\def\Ga{\Gamma}

\let\<\langle
\let \>\rangle

\newcommand{\rem}[1]{}

\newcommand{\de}{\delta}

\newcommand{\bGam}{\boldsymbol{\Gamma}}

\newcommand{\bxi}{\boldsymbol{\xi}}

\newcommand{\bs}{\mathbf{s}}

\newcommand{\bY}{\boldsymbol{Y}}

\newcommand{\bchi}{\boldsymbol{\chi}}

\newcommand{\pp}[2]{\frac{\partial #1}{\partial #2}}

\newcommand{\dede}[2]{\frac{\delta #1}{\delta #2}}

\newcommand{\om}{\omega}

\newcommand{\Om}{\Omega}

\newcommand{\mso}{\mathfrak{so}}

\newcommand{\todoDH}[1]{\vspace{5 mm}\par \noindent
\framebox{\begin{minipage}[c]{0.95 \textwidth}
\color{blue}\tt DH: #1 \end{minipage}}\vspace{5 mm}\par}

\newcommand{\todoVP}[1]{\vspace{5 mm}\par \noindent
\framebox{\begin{minipage}[c]{0.95 \textwidth}
\color{magenta}\tt VP: #1 \end{minipage}}\vspace{5 mm}\par}

\newcommand{\revision}[2]{#2} 

\begin{document}
\title[Non-holonomic systems with stochastic transport]{Dynamics of non-holonomic systems with stochastic transport }
\author{D. D. Holm and V. Putkaradze}

\date{\today}

\address{DDH: Department of Mathematics, Imperial College, London SW7 2AZ, UK. Email: d.holm@ic.ac.uk, \url{http://wwwf.imperial.ac.uk/~dholm/}}
\address{VP: Mathematical and Statistical Sciences, University of Alberta, Edmonton, T6G2G1, Canada. Email: putkarad@ualberta.ca, \url{http://www.mathstats.ualberta.ca}}

\begin{abstract}
This paper formulates a variational approach for treating observational uncertainty and/or computational model errors as stochastic transport in dynamical systems governed by action principles under nonholonomic constraints. For this purpose, we derive, analyze and numerically  study the example of an  unbalanced spherical ball rolling under gravity along a stochastic path. Our approach uses  the Hamilton-Pontryagin variational principle, constrained by a stochastic rolling condition, which we show is equivalent to the corresponding stochastic Lagrange-d'Alembert principle.  In the example of the rolling ball, the stochasticity represents uncertainty in the observation and/or error in the computational simulation of the angular velocity of rolling. 
The influence of the stochasticity on the deterministically conserved quantities is investigated both analytically and numerically.   Our approach applies to a wide variety of stochastic, nonholonomically constrained systems,  because it preserves the mathematical properties inherited from the variational principle.
\end{abstract}
\maketitle



\section{Introduction}
  The derivation and analysis of equations of motion for nonholonomic deterministic systems has a long history and remains a topic of active research \cite{Bloch2003}. A new set of challenges arises when such systems become stochastic.  We introduce stochasticity that represents uncertainty in the observations and/or simulations  of the angular velocity of rolling,  e.g., due to finite time steps between observations or computations.  For this example, we investigate the effects of this stochasticity in the angular velocity, in the presence of the nonholonomic rolling constraint. 
  
  Stochastic Hamiltonian systems were introduced and analyzed in the foundational work \cite{Bi1981}. These considerations were updated and recast in the language of geometric mechanics in \cite{LaOr2008}, inspiring new  considerations such as symmetry reduction and  Noether's theorem in the presence of  stochasticity. 
In this context, the papers \cite{Ho2010,HoRa2015} introduced mechanical systems subjected to stochastic forces while obeying nonholonomic constraints.  The physical background for  the systems considered there could be understood as the dynamics of a microscopic object bombarded by outside molecules, while preserving a nonholonomic constraint, such as rolling contact. This problem is highly non-trivial, as the reaction forces generated by the constraint require careful consideration. 

A different case of stochasticity arising in nonholonomic systems was considered by \cite{FGBPu2015}, where no random forces were acting on the system itself, but the constraint was stochastic. A physical realization of such system would be the motion of a deterministic rolling ball on a rough plane, or experiencing a random slippage at the contact point. It was shown that depending on the type of stochasticity in the constraint itself, the system can preserve some integrals of motion, and the energy is preserved in a general nonholonomic systems as long as the constraint remains homogeneous in velocities. This result contrasts  with \cite{HoRa2015}, where no integrals of motion were found to be preserved for general stochastic forces. The preservation of integrals of motion described in \cite{FGBPu2015} was due to the nature of the forces introduced by the stochastic constraints. 

In this paper, we consider another possible case of stochasticity in nonholonomic systems.  The noise we consider arises from, for example, errors in  observations of transport velocity for the mechanical systems. For example, imagine a rolling ball which is being recorded by a video camera. The measurement of the orientation and velocity of the ball will always be subject to errors, which will produce a deviation from the expected deterministic trajectory.   Unseen irregularities in the surfaces in contact may also cause intermittent changes in angular velocity, without violating the rolling constraint. We shall refer to this class of problems as \emph{stochastic dynamics with transport noise}.    Stochastic transport (ST) noise for systems was introduced in the context of fluid dynamics in \cite{Holm2015}. 
 For other recent investigations of  ST dynamics, we refer the reader to   \cite{CoGoHo2017, CrFlHo2017,Ho2017,ArGaHo2017,HoTy2016integrators,HoTy2016soliton,ArCaHo2016orbits,ArCaHo2016rigid,CrHoRa2016}.
In this paper, we investigate the effect ST has on nonholonomically constrained systems, \revision{R2Q7}{both analytically and numerically}. As it turns out, ST for nonholonomic systems affords an elegant and easy consideration of stochastic nonholonomic mechanics \revision{R2Q2}{for the case of rolling-ball type systems}, with the appropriately generalized applications of the Hamilton-Pontryagin principle and the Lagrange-d'Alembert principle.

\subsection*{Main content of the paper}
This paper formulates an approach for quantifying uncertainty in rolling motion by deriving, analysing and numerically simulating the equations for an unbalanced spherical ball \revision{R1Q1b}{with stochasticity caused by observation uncertainty. More precisely, the} stochasticity represents  uncertainty in the angular velocity at which the nonholonomic rolling constraint is imposed. The stochastic path is reconstructed from the solution of the dynamical system on which a noisy nonholonomic angular velocity constraint is imposed \revision{R2Q2}{on the motion of the group $SE(3)=SO(3) \times \mathbb{R}^3$, the group of motion for the rolling ball}. For this purpose, we apply  the Hamilton-Pontryagin variational principle \cite{Holm2008}, constrained by a stochastic rolling condition which represents error in the observation of the angular velocity of the rolling ball. The resulting stochastic dynamical system provides a method of quantifying uncertainty in measurements or numerical simulations of the rolling motion. To obtain the solution of the stochastic dynamical system, one first integrates the motion for angular momentum and solves for the body angular velocity and orientation of the spatial vertical direction, as seen in the body. This allows reconstruction of the time-dependent orientation, represented as a stochastic curve on the rotation group $SO(3)$. Finally, one applies the rolling constraint to obtain the position of the centre of mass of the rolling ball along the stochastic path. The examples treated are balanced and unbalanced rolling balls, as well as a balanced vertically rolling disk. 

\noindent
The paper proceeds as follows 
\begin{enumerate} 
\item Section~\ref{sec:general_principles} presents the derivation of stochastic evolution equations for a system with \revision{R1Q2a}{ noisy variational systems with nonholonomic constraints}, by using the Hamilton-Pontryagin and Lagrange-d'Alembert variational principles. \revision{R1Q2a}{The noise in the system models observation uncertainty (e.g., temporal resolution) rather than the effect of external random force.} These variational principles are applied to the case of unbalanced rolling ball, which is a classic example of a nonholonomic system. 
\item Section~\ref{sec:vertical_disk} studies the example of a vertically rolling disk, where analytical solutions for the dynamical quantities can be obtained in terms of Stratonovich integrals. \revision{R2Q7}{Numerical simulations of the rolling disk are also performed to illustrate the theoretical results.}
\item Section~\ref{sec:conservation_laws} computes the evolution equations for the analogues of the first integrals of deterministic rolling: energy, Jellet and Routh. We show that none of these classical integrals of motion are preserved for the stochastic case. We also study a particular case of Chaplygin sphere, when the center of mass coincides with the geometric center, and derive the evolution equations for the quantities that are conserved in the analogous deterministic case, but are not conserved in the stochastic case. 
\revision{R2Q7}{Numerical simulations of the rolling sphere are  performed to compute the variability of the energy, as well as that of the Jellet and Routh quantities and to illustrate the motion of the ball in space. }
\item Section~\ref{sec:summary} summarizes the paper and discusses other  problems treatable by our method. 
\end{enumerate} 
\revision{R2Q11}{We mention some nomenclature from the literature, as well. The term "Chaplygin's ball" refers to a (possibly inhomogeneous) sphere whose center of mass coincides with the geometric center. A rolling sphere whose center of mass does not lie at the geometric center is often called a "Chaplygin top" in the Russian literature, or a "Routh sphere" in the British literature.   }
\rem{ 
{\obeylines
Plan of the paper:
1. Introduction 
2. Nonholonomic stochastic variational principles 
2.1. An unbalanced ball, rolling stochastically without slipping 
Problem statement 
2.2. Hamilton-Pontryagin variational principle for a stochastically rolling ball 
2.3. Lagrange-d'Alembert variational principle for a stochastically rolling ball 
Preparation for Noether's theorem 
3. Conservation laws 
3.1. Cylindrically symmetric bodies, rolling along stochastic paths 
The Jellet integral. 
The Routh integral. 
4. Numerical Examples 
4.1. Stochastic rolling of Chaplygin's concentric sphere 
4.2. Circular disk rocking in a vertical plane
}
} 

\section{Nonholonomic stochastic variational principles} 
\label{sec:general_principles} 
\subsection{An unbalanced ball rolling with uncertain velocity}
\label{sec:roliing_ball_problem statement}

\subsection*{Problem statement}
\revision{R2Q8}{This section applies the  Hamilton--Pontryagin approach \cite{Holm2008} to a class of constrained action integrals which includes the motion of an unbalanced spherical ball rolling stochastically on a horizontal plane in the presence of gravity. For deterministic nonholonomic systems that are affine in velocity, this section verifies the equivalence of the Hamilton--Pontryagin principle and the more standard Lagrange--d'Alembert principle by direct computation \cite{Holm2008}. This equivalence can be also established for  the problems considered here, or, more generally, for problems formulated on  semidirect-product Lie groups, such as the rolling ball. We shall not discuss the equivalence between these two principles in general, as this topic is quite complex and is beyond the scope of this paper. We will use the Hamilton--Pontryagin principle in the remainder of the paper. } 

The stochasticity in this formulation models the uncertainty in the observation of the velocity of the rolling ball and leads to the reconstruction of the rolling path as a stochastic curve in the semidirect-product Lie group of Euclidean motions $SE(3) \simeq SO(3)\circledS\mathbb{R}^3$. As a space, the Euclidean group is $SE(3)\simeq SO(3)\times \mathbb{R}^3$. Consequently, the generalized velocities lying in the Lie algebra of the Euclidean group will be stochastic vectors in $\mathfrak{so}(3)\times \mathbb{R}^3$. These stochastic vectors will transport the configuration of the rolling ball along the stochastic path. 
The observations of the ball's rolling motion are described as transport by the Lie group action of $SE(3)$ that maps a generic point $P$ in the ball's reference coordinates to a point in space, $Q(t)$, at time $t$, according to
\[
Q(t)=g(t)P+x(t)
\,,  
\quad\hbox{where}\quad
g(t) \in SO(3)
\quad\hbox{and}\quad
 x(t)\in \mathbb{R}^3 \, ,
\] 
\revision{R1Q2d}{along a path of stochastic processes $(g(t),x(t))\in SE(3)$ parameterized by time, $t$.}
Here, $g(t) \in SO(3)$ represents the orientation of the ball, \revision{R3Q9}{and 
\[
x(t)=(x_1(t),x_2(t),x_3(t))\in \mathbb{R}^3
\]
is the projection of the point $P$, chosen to be the ball's center of mass.  We will choose the initial point $P$ to be} located in an equilibrium position below the ball's \revision{R2Q10}{geometric center} along the vertical direction in space, $e_3$. 

\begin{figure}[h]
\centering
\includegraphics[width=0.75\textwidth]{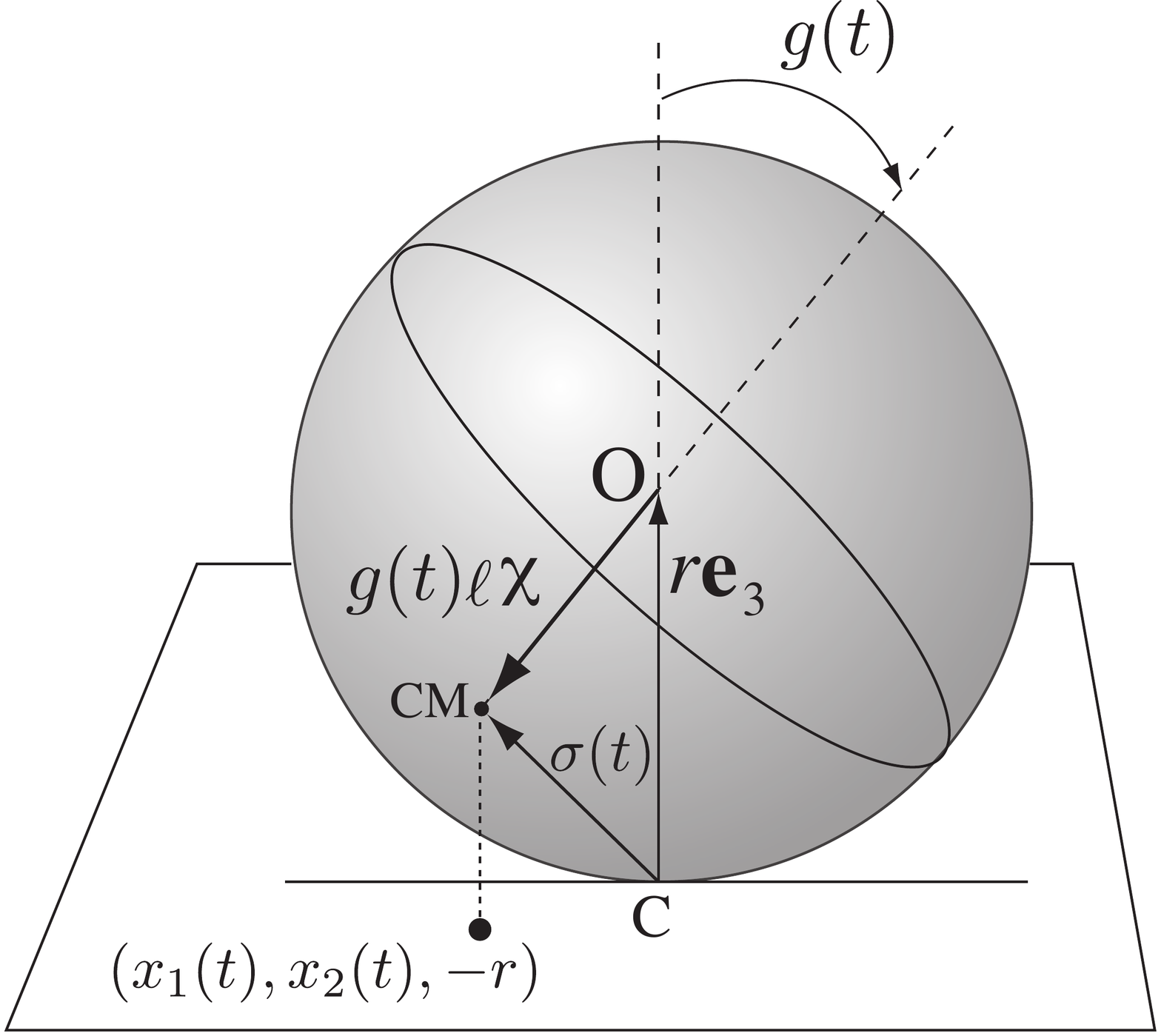}
\label{ball-fig}
\caption{\footnotesize
 The Chaplygin ball. The position $(x_1(t),x_2(t),x_3(t))$ is the position of the centre of mass, not the centre of the sphere. The spatial vector $\sigma(t)$ points from the contact point $\mathbf{C}$ to the centre of mass. The  projection of the centre of mass location onto the plane is  the point $(x_1(t),x_2(t),-r)$.} 

\end{figure}

{\bf Rolling constraint.}
The {\bfi spatial nonholonomic constraint distribution} $\mathcal{D}$ defining the rolling motion, when written in terms of a stochastic processes on the tangent bundle of $SE(3)$, is given by \cite{Holm2008}

\begin{align}
\begin{split}
{{\sf d}x}(t)
&={{\sf d}g}(t) {g}^{-1}(t) \sigma(t)
\\&
=:{{\sf d}g}(t) (r {g}^{-1}(t)e_3 + \ell \chi)
\\&
={{\sf d}g}(t)\big(r \Ga(t)+\ell\chi\big) =: {{\sf d}g}(t)\,s(t)
\,,
\end{split}
\label{spat-constraint}
\end{align}
where ${\sf d}$ denotes the stochastic evolution operator,
$r e_3$ is the spatial vertical vector from the point of contact $C$ on the plane to the centre of the ball, $\ell\chi$ is the vector displacement in the body pointing along the unit vector from the ball's \revision{R2Q10}{geometric center} to its center of mass, $\sigma(t)$ is their sum as spatial vectors, and $s(t)$ is their sum as body vectors. Figure \ref{ball-fig} sketches the configuration of the \emph{spatial vectors} at some time, $t$.

As introduced in \eqref{spat-constraint}, the time dependent vectors $s(t)$ and $\Ga(t)$ in the body reference frame are defined by
\begin{equation}
s := {g}^{-1}\sigma(t) =r \Ga + \ell\chi
\,,\quad\hbox{with}\quad
\Ga := {g}^{-1} e_3
\,.
\label{subst_Y}
\end{equation}


We decompose the stochastic quantities $g^{-1}{{\sf d}x}\in \mathbb{R}^3$ and $g^{-1}{{\sf d}g} \in \mathfrak{so}(3)$ into their drift $(dt)$ and noise $(dW^i)$ components (in the Stratonovich representation, denoted as $\circ \,dW^i$) following the spatial representation of the rolling constraint \eqref{spat-constraint} written in the body frame as
\begin{equation}
{g}^{-1} {{\sf d}x} = (g^{-1}{{\sf d}g}) s 
\,.
\label{body_constraint}
\end{equation}
Thus, we find the following stochastic rolling relations in the body representation, 

\begin{align}
\begin{split}
g^{-1}{{\sf d}g} &= \Omega\,{dt} + \sum_i \xi_i\circ dW^i(t) 
=: {\widetilde\Omega} \in  \mathfrak{so}(3)\simeq\mathbb{R}^3\,,
\\
g^{-1}{{\sf d}x} &= \Omega  s\,{dt} + \sum_i \xi_i  s\circ dW^i(t)
\\&=
Y\,{dt} + \sum_i \xi_i  s\circ dW^i(t)
=: \widetilde{Y} = \widetilde \Omega  s \in \mathbb{R}^3
\,.
\end{split}
\label{drift-stoch-decomps}
\end{align}
\revision{R2Q4}{This way of incorporating noise respects the rolling constraint, since $\widetilde{Y} = \widetilde \Omega  s$. }

%
%
%
\color{black} 
In \eqref{drift-stoch-decomps}, the quantities $\xi_i\in \mathfrak{so}(3)\simeq\mathbb{R}^3 $ with $i=1,2,\dots,({\dim}\,\mathfrak{so}(3)=3)$ comprise a set of \emph{fixed} Lie algebra elements, which may be identified with vectors in $\mathbb{R}^3$ by the familiar hat map, $(\,\cdot\,)\,\widehat{\ }: \mathfrak{so}(3) \to \mathbb{R}^3$, and $dW^i(t)$ denotes a set of independent Brownian motions. 
\revision{R2Q11}{The set $\xi_i$ need not span the whole space, and may contain more than three elements. Thus, the $\xi_i$ do not necessarily form a basis for $\mso(3)$, although they could be chosen to do so. }

Note that the stochastic decomposition in equations \eqref{drift-stoch-decomps} still satisfies the deterministic rolling condition. That is, 

\begin{align}
\widetilde{Y} - \widetilde \Omega\,s 
=
(Y-\Omega\,s)\,{dt} 
=0
\,.
\label{drift-stoch-decomps-roll}
\end{align}

\subsection{Hamilton--Pontryagin variational principle for a stochastically rolling ball}
We shall introduce the relations \eqref{spat-constraint} -- \eqref{drift-stoch-decomps} as constraint  equations for the stochastic motion of the rolling ball, as determined from the Hamilton--Pontryagin variational principle, $\delta S = 0$, applied to the following constrained stochastic action integral
\begin{align}
\begin{split}
S &=
 \int \bigg(
l(\Om, Y, \Ga ) 
+\
 \Big\langle\kappa\,,\,g^{-1}e_3-\Gamma \Big\rangle
 \bigg)dt
\\&\quad
+ \int \Big\langle\lambda\,,\,g^{-1}{{\sf d}x}- Ydt -   \sum_i \xi_i s{(\Gamma)} \circ dW^i(t)  \Big\rangle
\\&\quad
+ \int\Big\langle\Pi\,,\,g^{-1}{{\sf d}g} - \Omega\,{dt} - \sum_i \xi_i\circ dW^i(t) \Big\rangle
\,.\end{split}
\label{reduced-constrained-lag}
\end{align}
\revision{R1Q2e\\R1Q2f}{The last two integrals in \eqref{reduced-constrained-lag} include  Stratonovich stochastic  integrals, and $l(\Om,Y,\Ga)$ is the Lagrangian. } When the $\xi_i$ vanish, the action integral \eqref{reduced-constrained-lag} reduces to the deterministic case with the standard constraints for rolling without slipping, as discussed in textbooks, e.g., \cite{Holm2008}. 
\revision{R2Q5-6}{If $Y$ were absent and the middle line in equation \eqref{reduced-constrained-lag} were missing, then this problem would reduce to the stochastic heavy top, studied in \cite{ArCaHo2016rigid,ArCaHo2016orbits}.
}


\begin{theorem}[Hamilton--Pontryagin \ principle]
\label{unholy-rolling-motion-thm}$\,$\\
The stationarity condition for the nonholonomically constrained Hamilton--Pontryagin principle defined in equation (\ref{reduced-constrained-lag}) under the rolling constraint in body coordinates given in \eqref{body_constraint} by
\begin{equation}
\widetilde{Y} = (g^{-1}{{\sf d} x}) = (g^{-1}{{\sf d} g}) s = \widetilde{\Omega} s
\,,
\label{body-constraint}
\end{equation}
implies the following stochastic equation of motion,
\begin{equation}
\Big({\sf d} - {\rm ad}^*_{\widetilde\Omega}\Big)
\big(\Pi-\lambda\diamond s)
= \kappa \diamond\Gamma\,dt
- \lambda \diamond {\sf d}{s}
\,,
\label{UnholyHamPont-eqn}
\end{equation}
where $s=s(\Ga)=r \Ga + \ell\chi$ is the vector in the body directed from the point of rolling contact to the centre of mass. 

In addition, the Lagrange multipiers in the constrained stochastic action integral in (\ref{reduced-constrained-lag}) are given by 
\begin{align}
\Pi=\frac{\delta l}{\delta\Omega}
\,,\quad
\kappa\,dt=\frac{\delta l}{\delta \Ga}\,dt
\,{-\,r\sum_i \xi_i^T \lambda  \circ {d} W^i }
\,,\quad
\lambda=\frac{\delta l}{\delta Y}
\,.
\label{Lag-mults-def-theorem}
\end{align}
\end{theorem}

Before explaining the proof, we recall the definitions of the operations ${\rm ad}^*$ and $\diamond$ in the statement of Theorem \ref{unholy-rolling-motion-thm}, discussed in e.g. \cite{HoMaRa1998,Holm2008}.
\begin{remark}[The ${\rm ad}^*$ and $\diamond$
operations]\label{ad+diamond-defs} $\,$\\
The coadjoint action 
\[
{\rm ad}\,^*:\, \mathfrak{so}(3)^*\times\mathfrak{so}(3)\to\mathfrak{so}(3)^*
\]
and the diamond operation 
\[
\diamond:\,V^*\times V\to \mathfrak{so}(3)^*
\]
appearing in equation (\ref{UnholyHamPont-eqn}) are defined, respectively, as the duals of the Lie algebra adjoint action ${\rm ad}:\, \mathfrak{so}(3)\times\mathfrak{so}(3)\to\mathfrak{so}(3)$ and of its (left) action $ \mathfrak{so}(3)\times V\to V$ on the vector representation space $V=\mathbb{R}^3$ for the corresponding pairings, as follows. 

The operation ${\rm ad}\,^*$ is defined as the dual of the {\rm ad} operation,
\begin{equation}
\Big\langle {\rm ad}^*_{\widetilde\Omega}\,\Pi\,,\, \eta \Big\rangle
_{\mathfrak{so}(3)^*\times \mathfrak{so}(3)}
=
\Big\langle \Pi\,,\, {\rm ad}_{\widetilde\Omega}\, \eta \Big\rangle
_{\mathfrak{so}(3)^*\times \mathfrak{so}(3)}
\,,
\label{ad-star-def}
\end{equation}
with ${\widetilde\Omega}\in\mathfrak{so}(3)$, $\eta\in\mathfrak{so}(3)$, and $\Pi\in\mathfrak{so}(3)^*$.

\color{black} 
For the diamond operation $\diamond$, we define 
\begin{align}
\begin{split}
\Big\langle \kappa \diamond \Ga\,,\, \eta \Big\rangle
_{\mathfrak{so}(3)^*\times \mathfrak{so}(3)}
&=
\Big\langle \kappa\,,\, -\,\eta \Ga \Big\rangle_{V^*\times V}
\\ \hbox{and deduce}\quad&
=
\Big\langle -\,\eta^T\kappa\,,\, \Ga \Big\rangle_{V^*\times V}
=
\Big\langle -\,\Ga \diamond \kappa \,,\, \eta \Big\rangle
_{\mathfrak{so}(3)^*\times \mathfrak{so}(3)}
\,,
\end{split}
\label{diamond-def}
\end{align}
with  $\eta=-\eta^T\in\mathfrak{so}(3),\, \kappa\in V^*$ and $\Ga\in  V$. By its definition in \eqref{diamond-def}, the diamond operation for the (left) action $ \mathfrak{so}(3)\times V\to V$ is antisymmetric, i.e., $\kappa \diamond \Ga = -\,\Ga \diamond \kappa$.

\end{remark} 

\begin{proof} 
One evaluates the variational derivatives in the constrained Hamilton's principle \eqref{reduced-constrained-lag} from the definitions of the variables as 

\begin{align}
\begin{split}
\de\widetilde\Omega=
\de (g^{-1}{{\sf d}g}) &= {{\sf d} \eta} + {\rm ad}_{\widetilde\Omega}\,\eta
\,,\\
\de\Ga=
\de (g^{-1}e_3) &= -\, \eta (g^{-1}e_3) = -\, \eta \Gamma
\,,
\\
\de \widetilde Y=
\de (g^{-1}{{\sf d}x}) &= {\sf d}(\eta s) 
+ ( {\rm ad}_ {\widetilde\Omega}\,\eta)s
= \de \Big(\widetilde\Omega \,s(\Gamma)\Big)
\,.
\end{split}
\label{constr-var-deriv}
\end{align}
Here $\eta := g^{-1}\de g$ and the last equation is computed from \cite{Holm2008}
\begin{eqnarray}
\de (g^{-1}{{\sf d}x}) 
&=&
-\eta\, g^{-1}{{\sf d}x} + g^{-1}\de ({\sf d}x)
\nonumber
\\&=& -\, \eta\,{\widetilde\Omega}\, s + {\sf d}(g^{-1}\de x) + {\widetilde\Omega}\,\eta s
\,.
\label{crux}
\end{eqnarray}

\begin{remark}
In computing formulas (\ref{constr-var-deriv}), one must first take variations of the definitions, and only afterward evaluate the result on the constraint distribution defined by 
$\widetilde Y=g^{-1}{{\sf d}x}={\widetilde\Omega}\, s$  and 
\[
g^{-1}\de{x}=(g^{-1}\delta{g})(r {g}^{-1}e_3+\ell\chi)=\eta s,
\]
cf. equation (\ref{spat-constraint}). 
\end{remark}

Expanding the variations of the Hamilton--Pontryagin action integral (\ref{reduced-constrained-lag})  using relations (\ref{constr-var-deriv}) and then integrating by parts yields
\begin{eqnarray}
\de S \!\!\!&=&
\!\!\! \int_a^b\! 
 \left\langle \frac{\de l}{\de \Omega}-\Pi\,,\,\de\Omega\right\rangle{d} t
+ 
 \left\langle \frac{\de l}{\de \Gamma} {d} t 
 - \kappa{d} t\,{-\,r\sum_i \xi_i^T \lambda  \circ {d} W^i } \,,\,\de \Gamma\right\rangle 
+ 
 \left\langle \frac{\de l}{\de Y}-\lambda\,,\,\de Y\right\rangle{d} t
 \nonumber\\
&&\hskip-6pt
-\
\left\langle
\Big( {\sf d} - {\rm ad}^*_{\widetilde\Omega}\Big)
\big(\Pi-\lambda\diamond s\big)
- \kappa \diamond\Gamma\,dt
+ \lambda \diamond {{\sf d}s}
\,,\,\eta \!\right\rangle
\nonumber\\&&\hskip24pt +\  
\Big\langle
\big(\Pi-\lambda\diamond s\big), 
\eta
\Big\rangle
\Big|_a^b
\,.
\label{HamPrinc-ChaplyginTop}
\end{eqnarray}
The last entry in the integrand arises from varying in the group element $g$ using formulas (\ref{constr-var-deriv}) obtained from  relation (\ref{crux}). 
Stationarity $(\de S =0)$ for the class of action integrals $S$ in equation (\ref{reduced-constrained-lag}) for variations
$\eta$ that vanish at the endpoints now proves the formula  for the constrained equation of motion (\ref{UnholyHamPont-eqn}) in the statement of the theorem, while it also evaluates the Lagrange multipliers $\Pi$, $\kappa$ and $\lambda$ in terms of variational derivatives of the Lagrangian, as

\begin{align}
\Pi=\frac{\delta l}{\delta\Omega}
\,,\quad
\kappa\,dt=\frac{\delta l}{\delta \Ga}\,dt
\,{-\,r\sum_i \xi_i^T \lambda  \circ {d} W^i }
\,,\quad
\lambda=\frac{\delta l}{\delta Y}
\,.
\label{Lag-mults-def}
\end{align}
\end{proof}

\begin{remark}[Stochastic volatility of the noise]
\label{Stoch-vol-sec}

Theorem \ref{unholy-rolling-motion-thm} for the Hamilton--Pontryagin principle persists and its proof still proceeds along the same lines, even if further uncertainty is introduced, in the form of volatility in the amplitude of the stochastic processes in the reconstruction relations. In particular, Theorem \ref{unholy-rolling-motion-thm} and its proof persists modulo small modifications when the stochasticity in the previous reconstruction relations \eqref{drift-stoch-decomps} for $g^{-1}{{\sf d}x}\in \mathbb{R}^3$ and $g^{-1}{{\sf d}g} \in \mathfrak{so}(3)\simeq \mathbb{R}^3$ is taken (in the Stratonovich representation) to be

\begin{align}
\begin{split}
g^{-1}{{\sf d}g} &= \Omega\,{dt} + \sum_i \xi_i\circ dW^i(t) 
=: {\widetilde\Omega} \in  \mathfrak{so}(3)\,,
\\
\hbox{with}\quad 
{\sf d}\xi_i &= \alpha_i(t)\,dt + \beta_i \circ \, dW^i(t)\,,
\\ &\hspace{-1.7cm}\hbox{and}\quad 
g^{-1}{{\sf d}x} = Y\,{dt} + \sum_i \xi_i \bs\circ dW^i(t) 
=: \widetilde{Y} = \widetilde \Omega\,\bs \in \mathbb{R}^3
\,.
\end{split}
\label{drift-stoch-decomps-fun}
\end{align}
Here, the quantities $\xi_i\in \mathfrak{so}(3)\simeq \mathbb{R}^3$ with $i=1,2,\dots,({\dim}\,\mathfrak{so}(3)=3)$ comprise a set of Lie algebra elements undergoing their own stochastic processes, \revision{R1Q2i}{$\alpha_i(t)$ and $\beta_i(t)$ are, correspondingly, prescribed drift and diffusion terms that may depend on time, but not any of the dynamical variables}, and $dW^i(t)$ is a set of three independent Brownian motions.
\end{remark}

\begin{remark}[Explicit form of the motion equation]
Expanding out the equation of motion in \eqref{HamPrinc-ChaplyginTop} and using the definitions in \eqref{Lag-mults-def} yields 

\begin{align}
\Big( {\sf d} - {\rm ad}^*_{\widetilde\Omega}\Big)
\big(\Pi-\frac{\delta l}{\delta Y}\diamond s\big)
- \left(\frac{\delta l}{\delta \Ga}\,dt
\,{+\,r\sum_i \xi_i \frac{\delta l}{\delta Y}  \circ {d} W^i }\right) \diamond\Gamma\,dt
+ \frac{\delta l}{\delta Y} \diamond {{\sf d}s}
=
0
\,.
\label{HP-moteqn-explicit}
\end{align}
\end{remark}

\begin{remark}[Vector notation]\label{remark-vectors}
For $g\in SO(3)$ the equation of constrained motion (\ref{UnholyHamPont-eqn}) arising from stationarity $(\de S =0)$ of the action  in (\ref{HamPrinc-ChaplyginTop}) may be expressed in $\mathbb{R}^3$ vector notation via the hat map isomorphism $\mathfrak{so}(3)\leftrightarrow\mathbb{R}^3$, as

\begin{equation}
\Big({\sf d} + \boldsymbol{\widetilde\Omega\times}\Big)
\big(\boldsymbol{\Pi-\lambda \times s})
= \boldsymbol{\kappa} \times\boldsymbol{\Gamma}\,dt
 - \boldsymbol{\lambda} \times {\sf d}\boldsymbol{s}
\,.
\label{UnholyHamPont-vectoreqn}
\end{equation}
These equations are completed by the formulas\\ 
\[
\boldsymbol{\Pi} =\dede{l}{ \bOm}
\,,\quad 
\boldsymbol{\kappa} {d} t =\dede{l}{ \boldsymbol{\Gamma}} {d} t
{\,-\,r \boldsymbol{\lambda} \times \sum_i \bxi_i  \circ {d} W^i }
\, , 
\quad 
\boldsymbol{\lambda} =\dede{l}{  \boldsymbol{Y}} 
\,,
\]
\[
d\boldsymbol{{\Ga}=-\,\boldsymbol{\widetilde\Omega\times} \Ga}
\,,\quad
\boldsymbol{\widetilde\Omega} := 
\boldsymbol{\Omega}\,{dt} + \sum_i \boldsymbol{\xi}_i\circ dW^i(t) 
\,,
\]
with ${\sf d}\boldsymbol{s}$ computed from $\bs=\bs(\bGam)= r \bGam + \ell \bchi$,
as ${\sf d}\boldsymbol{s} = r {\sf d}\bGam = - r \,\boldsymbol{\widetilde\Omega}\times \bGam$.

\end{remark}

\begin{remark}[Standard form in vector notation]\label{Vector-notation-rem}
So that our vector notation agrees with that of previous works on the subject of the rolling ball, in what follows, we will assume that the reduced Lagrangian in Equation (\ref{reduced-constrained-lag}) for the \revision{R2Q11}{Chaplygin top (Routh sphere)} in body coordinates is given by \cite{Holm2008}
\begin{equation}
l(\boldsymbol{\Om, Y, \Ga} ) 
=  \frac{1}{2} \boldsymbol{\Om\cdot} \mathbb{I} \boldsymbol{\Om}
+ \frac{m}{2} | \boldsymbol{Y} |^2 
- m \gamma \ell\,  \boldsymbol{\Ga \cdot \chi} 
\,.
\label{reduced-lag-vec}
\end{equation}
This is the sum of the kinetic energies due to rotation and translation, minus the potential energy of gravity. 
One then evaluates its vector-valued variational relations as
\begin{eqnarray}
\boldsymbol{\Pi} \!\!\!&=&\!\!\!  \frac{\de l}{\de \boldsymbol{\Omega}}
= \mathbb{I} \boldsymbol{\Omega}
\,,\quad
{
\boldsymbol{\kappa} \mbox{d}t =-\, m g\ell\boldsymbol{\chi}  \mbox{d}t
-\,r \boldsymbol{\lambda} \times \sum_i \bxi_i  \circ {d} W^i
}
\,,\quad
\boldsymbol{\lambda} = \frac{\de l}{\de \boldsymbol{Y}} 
= m \boldsymbol{Y}
\,.
\label{reduced-lag-varquants}
\end{eqnarray}
Thus, equation \eqref{UnholyHamPont-vectoreqn} is written explicitly in vector notation in the form 
\begin{equation} 
\label{Noisy_Ball_explicit1} 
\begin{aligned} 
\Big({\sf d} \ +\  \widetilde{\bOm} &\times \Big)  \left( \mathbb{I} \bOm + \bs \times m \bY\right) 
\\ &= 
mg\ell \bGam \times \bchi \,{d} t  {\,+\,}m \bY \times (\widetilde{\bOm} \times r \bGam) 
+
r {\bGam \times 
\left(   m \bY \times \sum_i \bxi_i  \circ {d} W^i 
\right) 
},
\end{aligned} 
\end{equation} 
where $\bY=\bOm \times \bs$.
\end{remark}

\revision{R1Q2j}{{\bf Standard Stochastic Differential Equation (SDE) form.} }
It is also useful to rewrite \eqref{Noisy_Ball_explicit1} in the standard SDE form, by separating noise and drift terms. For this, we separate $\widetilde{\bOm} = \bOm \mbox{d} t + \sum_i \bxi_i \circ \mbox{d} W^i$ and write  
\[ 
\begin{aligned} 
\Big({\sf d} \ +\  \bOm \mbox{d} t  &\times \Big)  \left( \mathbb{I} \bOm + \bs \times m \bY\right) 
\\ &= 
mg\ell \bGam \times \bchi \,{d} t  {+}m \bY \times (\bOm \times r \bGam) \mbox{d} t \
+
r \bGam \times 
\left(   m \bY \times \sum_i \bxi_i  \circ {d} W^i 
\right) 
\\
& \quad 
+
m \bY \times \left( \sum_i \bxi_i \circ \mbox{d} W^i   \times r \bGam \right) 
- 
\sum_i \bxi_i \circ \mbox{d} W^i  \times   \left( \mathbb{I} \bOm + \bs \times m \bY\right) 
\\ 
& = 
mg\ell \bGam \times \bchi \,{d} t  {+}m \bY \times (\bOm \times r \bGam) \mbox{d} t \
\\
& \quad 
-
\bxi_i \circ \mbox{d} W^i   \times \left(  m \bY \times  r \bGam \right) 
-\sum_i \bxi_i \circ \mbox{d} W^i \times \left( \mathbb{I} \bOm + \bs \times m \bY\right) 
\\
& = 
mg\ell \bGam \times \bchi \,{d} t  {+}m \bY \times (\bOm \times r \bGam) \mbox{d} t 
- 
\sum_i \bxi_i \circ \mbox{d} W^i   \times \left( \mathbb{I} \bOm 
+  \ell \bchi \times m \bY\right) 
\end{aligned} 
\] 
Then, the explicit version of \eqref{Noisy_Ball_explicit1} separating drift and noise term is obtained by setting $\bY = \bOm \times \bs$: 
\begin{equation} 
\begin{aligned} 
 \Big({\sf d} &\ +\  \bOm \mbox{d} t  \times \Big)    \left( \mathbb{I} \bOm + \bs \times m \big( \bOm \times \bs \big) \right) 
 =mg\ell  \bGam \times \bchi \,{d} t  
\\
& 
{+}m  \big( \bOm \times \bs \big)  \times (\bOm \times r \bGam) \mbox{d} t \
- 
\sum_i \bxi_i \circ \mbox{d} W^i  \times  \left( \mathbb{I} \bOm + \ell \bchi  \times m  \big( \bOm \times \bs \big) \right) 
,\end{aligned} 
\label{Noisy_Ball_explicit2} 
\end{equation} 
again with ${\sf d}\boldsymbol{s}$ computed from 
$\bs=\bs(\bGam)= r \bGam + \ell \bchi$,
as ${\sf d}\boldsymbol{s} = r {\sf d}\bGam 
= - r \,\boldsymbol{\widetilde\Omega}\times \bGam$.
\revision{R1Q2n}{For the purpose of plotting trajectories of the geometric center ${\bf x}_{\mbox{gc}}$ of the ball later, the following formula will be useful
\begin{equation} 
\label{xgc_eq}
{\sf d} {\bf x}_{\mbox{gc}} 
= \Lambda  \big( \bOm \mbox{d} t + \sum_i \bxi_i \circ\mbox{d} W^i \big) \times r \mathbf{e}_3 = \Lambda \widetilde{\bOm} \times r \mathbf{e}_3 \, , 
\end{equation} 
where $\mathbf{e}_3$ is the fixed vector in vertical direction in the spatial frame. One can see from \eqref{xgc_eq} that the vertical coordinate of the geometric center ${\bf x}_{\mbox{gc}}  \cdot \mathbf{e}_3$ is preserved exactly, as expected. 
}

\noindent 
\revision{R1Q2k}{{\bf It\^{o} form of \eqref{Noisy_Ball_explicit2}.}
It is also useful to write \eqref{Noisy_Ball_explicit2} in It\^{o} form as this formulation is frequently used for numerical solutions of SDEs 
\cite{KlPl1992}. For notational convenience, we write this equation in the following compact form: 
\begin{equation} 
\left\{ 
\begin{aligned}
{\sf d} \bOm &= \mathbf{a}_{\bOm} \mbox{d} t + \sum_{i=1}^n \mathbf{b}_{\bOm}^i \circ \mbox{d} W^i\, , \quad 
 \mathbf{b}_{\bOm}^i:=\mathbb{I}^{-1} \Big( \bxi_i \times \left( \mathbb{I} \bOm + \ell \bchi \times m \big( \bOm\times \bs(\bGam) \big)  \right) \Big) 
\\ 
{\sf d} \bGam&=\mathbf{a}_{\bGam} \mbox{d} t + \sum_{i=1}^n  \mathbf{b}_{\bGam}^i \circ \mbox{d} W^i
\, , \quad 
\mathbf{b}_{\bGam}^i:=\bxi_i(t) \, . 
\label{Short_form_SDE}
\end{aligned} 
\right. 
\end{equation}
Notice that the drift $\mathbf{b}_{\bGam}^i$ does not depend on the stochastic variables $\bOm$ and $\bGam$. The Stratonovich-to-It\^{o} conversion formula then yields 
\begin{equation}
\left\{ 
\begin{aligned}
{\sf d} \bOm &= 
\mathbf{a}_{\bOm, \mbox{It\^{o} }} \mbox{d} t + \sum_{i=1}^n \mathbf{b}_{\bOm}^i \mbox{d} W^i\, , \quad 
\mathbf{a}_{\bOm, \mbox{It\^{o} }} ^i=\mathbf{a}_{\bOm}
-\frac{1}{2} \sum_{i=1}^n 
\Big(\mathbf{b}_{\bOm}^i \cdot \pp{\mathbf{b}_{\bOm}^i }{\bOm} + 
\bxi^i \cdot \pp{\mathbf{b}_{\bOm}^i }{\bGam}\Big)
\\ 
{\sf d} \bGam&=\mathbf{a}_{\bGam,\mbox{It\^{o}}} \mbox{d} t + \sum_{i=1}^n  \mathbf{b}_{\bGam}^i  \mbox{d} W^i
\, , \quad 
\mathbf{a}_{\bGam, \mbox{It\^{o}} }^i:=\mathbf{a}_{\bGam }^i \, , 
\end{aligned} 
\right. 
\label{drift_conversion}
\end{equation} 
where one  uses the dependence of $\mathbf{b}_{\bGam}(\bOm,\bGam,t)$ on $\bOm$ and $\bGam$ given by  
\eqref{Short_form_SDE} to compute the derivatives in the new drift terms $\mathbf{a}_{\bGam, \mbox{It\^{o}}}^i$. 
}

\subsection{Lagrange-d'Alembert variational principle for a stochastically rolling ball}

This section will show that the Lagrange-d'Alembert principle recovers precisely the same equation \eqref{HP-moteqn-explicit} for a stochastically rolling ball, as was obtained from the Hamilton-Pontryagin approach in the previous section. 

The Lagrange--d'Alembert principle on the tangent space $TG$ of a Lie group $G$ acting on a vector space $V$ with equations of motion on $T^*(G\times V)$ is equivalent to a constrained variational principle on $T(G\times V)$ with Euler--Poincar\'e equations on $\mathfrak{g}^*\times V^*$ \cite{Bloch2003,Holm2008}.  This equivalence arises because the integrand for the Lagrange--d'Alembert principle for $L$ on $T(G\times V)$ in the stationary principle is equal to the integrand of the reduced Lagrangian $l$ on $\mathfrak{g}\times V$.  We must compute what the variations on the group $G$ imply on the reduced space $\mathfrak{g}\times V$.  Define $\eta = g^{-1} \de g$.  As for the pure Euler--Poincar\'e theory with left-invariant Lagrangians, the proof of the variational  formula for $\Omega = g^{-1}{\sf d}g$ expressing $\de \Omega$ in terms of $\eta$ proceeds by direct computation, along the lines of \eqref{constr-var-deriv} and \eqref{crux}. 

Upon rearranging the stochastic rolling relations \eqref{drift-stoch-decomps} in the body representation, we find

\begin{align}
\begin{split}
\Omega\,{dt}  &=  g^{-1}{{\sf d}g} - \sum_i \xi_i\circ dW^i(t)  
=: {\widetilde\Omega} - \sum_i \xi_i\circ dW^i(t) \,,
\\
Y\,{dt} &= g^{-1}{{\sf d}x} - \sum_i \xi_i  s(\Ga)\circ dW^i(t) = \Omega  s\,{dt} 
\\
&=: \widetilde{Y} - \sum_i \xi_i  s(\Ga)(\Ga)\circ dW^i(t) 
=: {\widetilde\Omega}s - \sum_i \xi_i  s(\Ga)\circ dW^i(t)  
\,.
\end{split}
\label{solving-drift-stoch-decomps}
\end{align}
Taking variations using \eqref{constr-var-deriv} and \eqref{crux}, and substituting 
$s(\Ga) = r\Ga + \ell\chi$ yields

\begin{align}
\begin{split}
\de \Omega\,{dt}  &=  \de(g^{-1}{{\sf d}g}) 
= {\sf d}\eta +{\rm ad}_{\widetilde\Omega}\eta
\,,
\\
\de Y\,{dt} &= ({\sf d}\eta +{\rm ad}_{\widetilde\Omega}\eta)s + \eta {\sf d}s
- r\Big(\sum_i \xi_i  \circ dW^i(t)\Big)\de \Ga
\,,\\
\de \Ga &= -\eta \Ga
\,.
\end{split}
\label{LdA-variations}
\end{align}
After this preparation, the nonholonomic EP equation finally emerges from a \emph{direct} computation of stationarity 
of the variation, $\delta{S}$, of the action $S=\int_a^b l(\Omega, Y,\Ga)\,dt$:

\begin{align}
\begin{split} 
0 = \de {S}
&= 
\int_a^b \left\langle \frac{\partial l}{\partial \Omega} , \de \Omega \right\rangle 
+ \left\langle \frac{\partial l}{\partial Y} , \de Y \right\rangle
+
 \left\langle \frac{\partial l}{\partial \Ga} , \de \Ga \right\rangle dt 
 \\
\rem{ 
&= 
\int_a^b \left\langle \frac{\partial l}{\partial \Omega} , \d{\eta} 
+ {\rm ad}_{\widetilde\Omega} \eta \right\rangle 
+  \left\langle \frac{\partial l}{\partial \Ga} , -\eta  \Ga \right\rangle \,dt 
\\&\hspace{7mm}
+ \left\langle \frac{\partial l}{\partial Y} , ({\sf d}{\eta}  
+ {\rm ad}_{\widetilde\Omega}\eta)  s 
+ \eta   {\sf d}{s} 
+ r\Big(\sum_i \xi_i  \circ dW^i(t)\Big) , \eta\Ga
\right\rangle 
\\
}
&= 
\int_a^b \left\langle 
\Big(-{\d} + {\rm ad}^*_{\widetilde\Omega}\Big)
\frac{\partial {l} }{\partial \Omega} , \eta \right\rangle 
+ 
 \left\langle 
 \frac{\partial l}{\partial \Ga}\diamond \Ga , \eta  
\right\rangle \,dt 
-
\left\langle 
 \frac{\partial {l}}{\partial Y} \diamond {\sf d}s  , \eta
\right\rangle
\\&\hspace{4mm}
+ \left\langle 
\Big(-{\d} + {\rm ad}^*_{\widetilde\Omega}\Big) 
\Big(s \diamond \frac{\partial {l}}{\partial Y}\Big) , \eta
\right\rangle 
+ 
\left\langle
\bigg(r\Big(\sum_i \xi_i  \circ dW^i(t)\Big)\frac{\partial {l}}{\partial Y}\bigg)\diamond \Ga , \eta
\right\rangle 
\\&
\hskip24pt +\  
\bigg\langle
\Big(\frac{\partial l}{\partial \Omega} + s \diamond \frac{\partial {l}}{\partial Y}\Big), 
\eta
\bigg\rangle
\bigg|_a^b
\end{split} 
\label{Symm-Var}
\end{align}
Thus, for variations that vanish at the endpoints, this calculation for the Euler-Poincar\'e reduction of the Lagrange-d'Alembert principle recovers precisely equation \eqref{HP-moteqn-explicit}, which was obtained from the Hamilton-Pontryagin approach. Hence, we have the following.

\begin{theorem}[H-P v L-d'A equivalence]
The nonholonomic motion equations obtained as extremal conditions for the Lagrange-d'Alembert variational principle in Euler-Poincar\'e form after left Lie group reduction are equivalent to those obtained from the corresponding Hamilton-Pontryagin variational principle.
\end{theorem}

\begin{figure}[h!]
\centering
\includegraphics[height=0.5\textwidth]{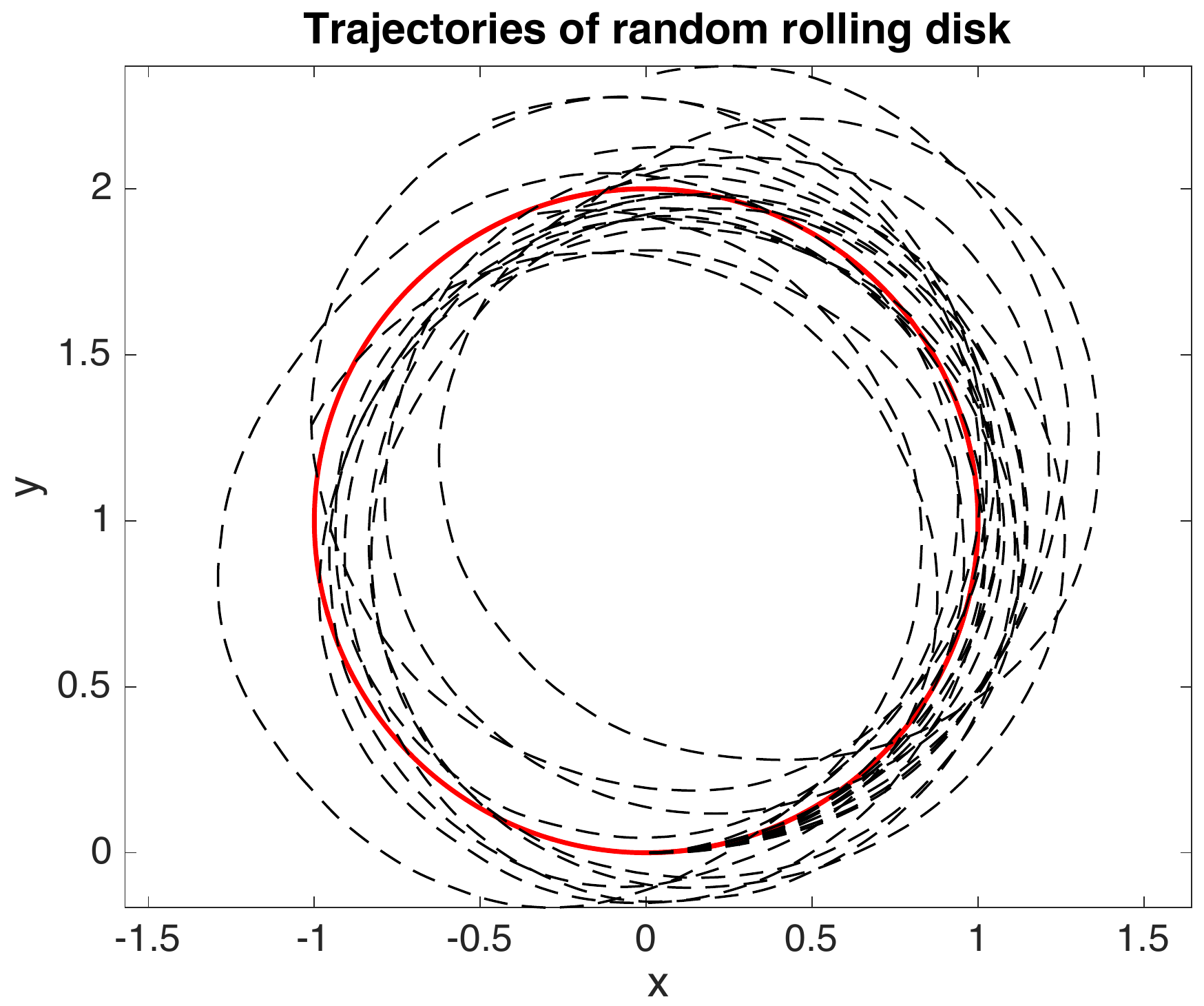} 
\caption{
\label{fig:Sim_Results_Disk} 
Trajectories of the disk's center obtained by numerically solving \eqref{rolling_disk_theta+phi} and \eqref{constraints_disk}, with $\xi_1 =\xi_2 = 0.1$ for 10 realizations of noise (dashed thin lines). The trajectory for the non-noisy case $\xi_1=\xi_2=0$ is also presented with a solid red line. All trajectories start at $(x_0,y_0)=(0,0)$ at $t=0$.  }  
\end{figure}

\revision{R2Q3}{\section{Analytically solvable case: the rolling vertical disk} }
\label{sec:vertical_disk}
In order to illustrate our methods, we present the even simpler case of a vertical rolling disk, where the solution can be written explicitly in terms of the stochastic integrals. We consider a flat, uniform disk of mass $m$, radius $r$ and moment of inertia taken about the rotation normal to the flat part of the disk being $I$ and any axis lying in the plane of the disk being $J$. The configuration space for the vertically rolling disk consists of 4 variables: $(x,y,\phi,\theta)\in \mathbb{R}^2\times S^1\times S^1$. In addition, two coordinates of the center of the disk projected onto the plane $(x,y)$, the configuration is specified by the angle of rotation $\phi \in S^1$ with respect to a spatially vertical axis in the plane of rolling, and the angle of rotation about the axis of symmetry $\theta\in S^1$. 
We define the angular velocities $\omega$ and $\nu$ and \revision{R1Q2m}{noise intensities} $\xi_1$ and $\xi_2$ by 
\begin{equation} 
{\sf d}\theta = \omega dt + \xi_1\circ dW_1(t) \, , 
\quad 
{\sf d}\phi = \nu dt + \xi_2\circ dW_2(t) \, . 
\label{drift_theta_phi}
\end{equation} 
We shall consider $\xi_{1,2}$ to be constants, for simplicity in what follows. 
The rolling constraints are nonholonomic and require that the disk is moving tangent to its sharp edge without slipping. These constraints are written  as 
\begin{equation} 
{\sf d} x = R \omega \cos \phi \,{d} t 
\, , \quad 
{\sf d} y = R \omega  \sin \phi \,{d} t \, . 
\label{constraints_disk}
\end{equation} 
The variational principle for the deterministic rolling vertical disk is explained in \cite{BlKrMaMy1996} pp 19-20, see also \cite{Bloch2003} pp.238-244.  In our case, the corresponding action integral for the stochastic constrained nonholonomic variational principle $0=\delta S$ is written via the Hamilton-Pontryagin principle as 
\begin{align}
\begin{split}
S &= \int \left(\frac{m}{2} (u^2+v^2) + \frac12 I\omega^2 
+ \frac12 J\nu^2  \right)dt
+ \Big( p_1 ({\sf d}x - u dt) + p_2 ({\sf d}y - v dt)\Big)
\\&\qquad
+ \Big( \mu_1 (u - R\om \cos\phi) + \mu_2 (v - R\om \sin\phi) \Big)dt
\\&\qquad
+ \Big( \pi_1 \big({\sf d}\theta - \om dt - \xi_1\circ dW_1(t)\big) 
+ \pi_2 \big({\sf d}\phi - \nu dt - \xi_2\circ dW_2(t)\big)\Big),
\end{split}
\label{rolling_disk_action} 
\end{align}
where the Lagrange multipliers $(\mu_1,\mu_2)$ impose constraints \eqref{constraints_disk}; $(p_1,p_2)$ define the velocities $(u,v)$; and $(\pi_1,\pi_2)$ introduce stochasticity into the angular motion. The variations are given by
\begin{align}
\begin{split}
\de x \,\&\, \de y  &:\  {\sf d}p_1 = 0 \, , \quad {\sf d}p_2 = 0 \,,
\\
\de u \,\&\,  \de v &:\ mu + \mu_1 - p_1 = 0 \, , mv + \mu_2 - p_2 = 0 \,,
\\
\de \theta \,\&\,  \de \phi &:\  -\,{\sf d}\pi_1 = 0 \, , \quad R\om (\mu_1\sin\phi - \mu_2\cos\phi)dt - {\sf d}\pi_2 = 0\,,
\\
\de \om \,\&\,  \de \nu &:\  I\om - \mu_1 R \cos\phi - \mu_2 R \sin\phi - \pi_1 = 0 \, , \quad  J\nu - \pi_2 = 0\,.
\end{split}
\label{rolling_disk_var} 
\end{align}
The first equation of \eqref{rolling_disk_var} implies that $p_1$ and $p_2$ are constants, and we will set them to be zero in what follows. 
The second equation of \eqref{rolling_disk_var} yields equation for $\mu_1$ and $\mu_2$ 

\begin{align}
\mu_1 &=  - mu   = - mR\om \cos\phi  
\, , \qquad 
\mu_2 = - mv  = -mR\om \sin\phi  \,.
\label{rolling_disk_mu1+mu2} 
\end{align}
After solving for $\pi_1$ and $\pi_2$,  the third and fourth equation of \eqref{rolling_disk_var} yield 

\begin{align}
\begin{split}
(I+mr^2) {\sf d}\om = 0 \, \quad J {\sf d} \nu =0 \, , \quad \Rightarrow \quad \omega=\omega_0 \, , \quad \nu =\nu_0 \, . 
\end{split}
\label{rolling_disk_eqns1} 
\end{align}
Finally, upon inserting \eqref{rolling_disk_mu1+mu2} into \eqref{rolling_disk_eqns1} to eliminate $\mu_1$ and $\mu_2$, then taking the stochastic time differential on the right side of the first equation of \eqref{rolling_disk_eqns1} and using the stochastic constraints for ${\sf d}\theta$ and ${\sf d}\phi$, we obtain 

\begin{align}
\begin{split}
{\sf d}\theta - \om dt - \xi_1\circ dW_1(t) &= 0  \quad \Rightarrow \quad \theta=\theta_0 + \omega_0 t + \int_0^t  \xi_1  \circ dW_1(\tau) 
\,,\\
{\sf d}\phi - \nu dt - \xi_2\circ dW_2(t) &= 0 \quad \Rightarrow \quad \phi=\phi + \nu_0 t + \int_0^t  \xi_2  \circ dW_2(\tau) 
\,.
\end{split}
\label{rolling_disk_theta+phi} 
\end{align}
Since $\xi_1$ and $\xi_2$ are constants, there is no distinction between Stratonovich and It\^o noise, so the distribution function for the shifted variables $\theta-\omega_0 t$ and $\phi-\nu_0 t$ tends to the uniform distribution on $(0,2 \pi)$. 
Finally, the trajectory of the disk on the plane is given by integrating the constraint equations $(x,y)$ given by \eqref{constraints_disk}, 
with $\phi(x)$ given by \eqref{rolling_disk_theta+phi}. Clearly, only the component $\xi_2$ of the noise contributes to the spatial trajectory. We present the results of simulations of equations \eqref{rolling_disk_theta+phi} and \eqref{constraints_disk} in  Figure~\ref{fig:Sim_Results_Disk}. All trajectories start at the origin at $t=0$, with $\xi_1=\xi_2=0.1$, and $\omega_0=\nu_0=1$. The trajectories initially stay close to the circle, which is the exact solution in the deterministic case, also presented in red in the Figure. As time proceeds, the solution deviates further from the deterministic solution. 

\section{Conservation laws and particular cases }
\label{sec:conservation_laws}
\subsection{Conservation laws for the Routh sphere}

Let us follow the classical results for the Routh sphere (Chaplygin ball), and investigate the preservation of integrals of motion.  Mathematically, this particular case is obtained when two moments of inertia are equal, which we take to be $I_1=I_2$, and the axis of the third moment of inertia coincides with the  direction $\bchi$. In what follows, we shall assume cylindrical symmetry of the ball $\mathbb{I}={\rm diag}(I_1,I_1,I_3)$, \emph{i.e.} $I_1=I_2$, and take the center of mass to be offset from the geometric center along the $\mathbf{E}_3$ direction, so that $\bchi =\mathbf{E}_3$. It is known that, in this case \cite{Bloch2003,Holm2008}, the deterministic dynamics preserves three integrals of motion: energy, Jellet and Routh, whose precise expressions will  be defined immediately below. 

\subsection*{Energy.}
Let us first consider the (full) energy defined as 
\begin{equation} 
\label{energy} 
E=\frac{1}{2} \mathbb{I} \bOm \cdot \bOm + \frac{1}{2} m |\bY|^2 + m g l \bGam \cdot \bchi. 
\end{equation} 
Taking a scalar product of the equation of motion \eqref{Noisy_Ball_explicit2} with the angular velocity $\bOm$, we notice that the stochastic evolutionary derivative or $E$ given by \eqref{energy} can be formulated as 
\begin{equation} 
\label{dE0} 
\hspace{-2mm}
\mbox{d} E= \sum_i \bxi_i \!\circ\! \mbox{d} W^i \!\cdot\! \left( m g l  \bGam \times \bchi + \mathbb{I} \bOm \times \bOm + r \bGam \times \big( m \bY \times \bOm \big) - \bOm \times \big( \ell \bchi \times m \bY \big)  \right) 
 \, . 
\end{equation} 
Thus, in general, energy is not conserved in stochastic rolling. 

\subsection*{The Jellet integral.}
Let us turn our attention to Jellet integral $J  = \mathbf{M} \cdot \bs$, where $\mathbf{M}:= \mathbb{I} \bOm + \bs \times \bY$, which is also conserved for the deterministic case. In Appendix~\ref{app:Jellet_Routh} we derive that ${\sf d} J \neq 0$ as  
\begin{equation} 
\label{dJ_expression2} 
\begin{aligned} 
\mbox{d} J & 
=\sum_i \bxi_i \circ \mbox{d} W^i \cdot \left( 
\ell \bchi \times 
\mathbb{I} \bOm + \ell \bchi \times \big( \ell \bchi \times m \bY \big) + \big(r \bGam \times m \bY \big) \times r \bGam 
\big) 
\right) 
 \,.
\end{aligned} 
\end{equation} 
Hence, the Jellet integral is \emph{not} conserved in the stochastic rolling of a  ball with axisymmetric mass distribution. 

\subsection*{The Routh integral.}
Let us now turn our attention to the derivation of the Routh integral, which in our variables can be written as $R=\Omega_3 \sqrt{I_1 I_3 + m \mathbb{I} \bs \cdot \bs}$. The derivation of the Routh integral is, in our opinion, rather technical and non-intuitive. In Appendix~\ref{app:Jellet_Routh} we present the derivation for the evolution of this quantity in the stochastic case and derive the following equation: 
\begin{equation} 
\label{Routh_cons2} 
\begin{aligned} 
\frac{1}{2} {\sf d} R^2 &= - \frac{s_3 \Omega_3}{I_1}{\sf d} J 
  \\ 
&+
\sum_i \bxi_i \circ {\sf d} W^i \cdot  
\left( \Omega_3  \bchi \times  \Big[
 \left( \mathbb{I} \bOm + \ell \bchi \times m \bY \right) - 
m \bs \frac{I_1 \ell + (I_3-I_1) s_3}{I_1} 
\Big] \right) ,
\end{aligned} 
\end{equation} 
where the stochastic evolution of Jellet integral ${\sf d} J$ is given by \eqref{dJ_expression2}. Hence, even in the case when \revision{R2Q14}{all $\bxi_i \parallel \bchi$}, the Jellet integral is \emph{not} conserved in the stochastic rolling of a  ball with axisymmetric mass distribution, so the Routh integral is also \emph{not} conserved. 

\begin{figure}[h!]
\centering
\includegraphics[width=0.8\textwidth]{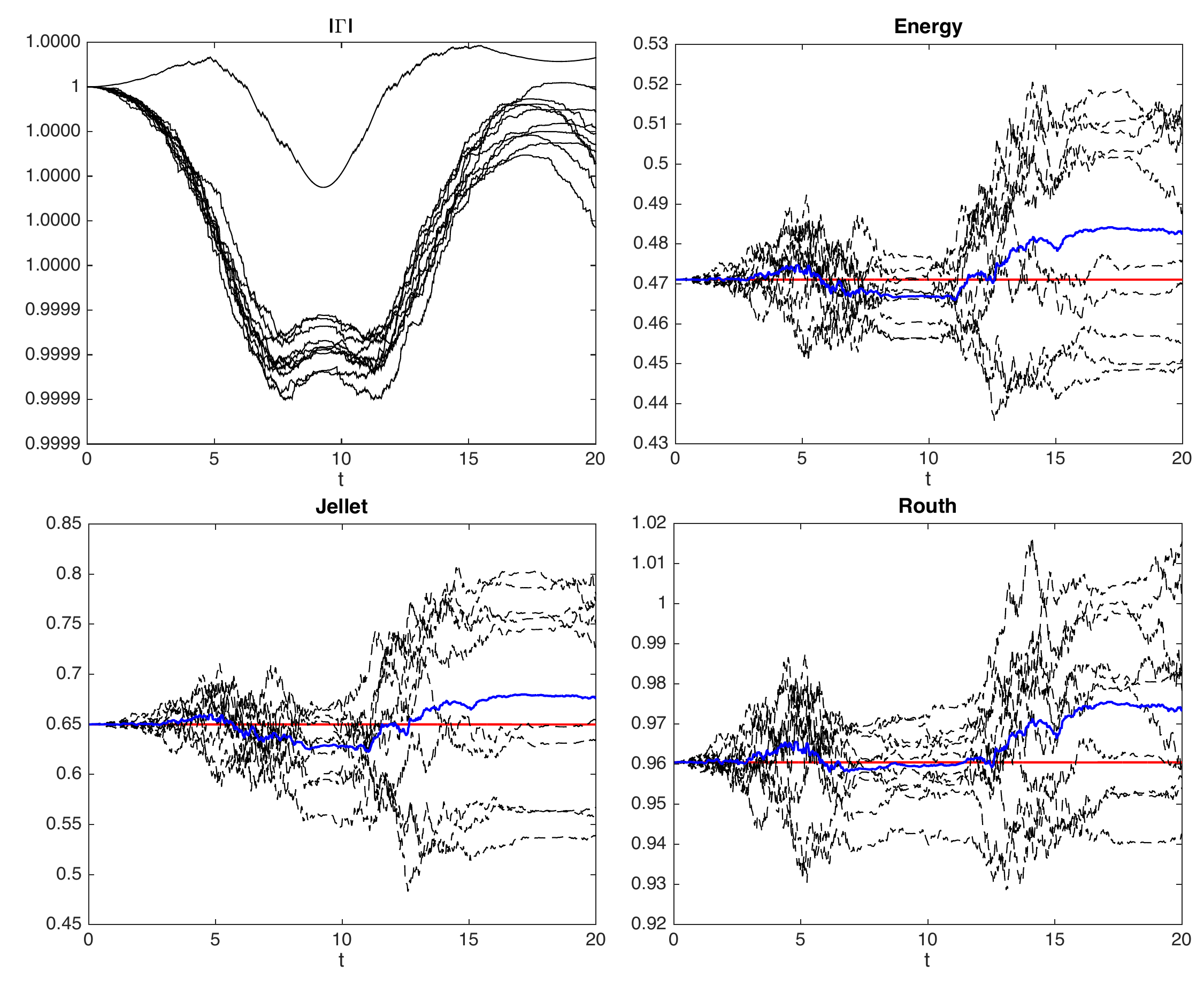} 
\caption{
\label{fig:Sim_Results} 
Results of simulations with $\bxi = 0.1 \bchi$, and time step $\Delta t=0.02$. Dashed black lines: different noise realizations. Blue solid line: mean values of different noise realizations. Red line: no noise (deterministic system).   Also included is the result of the simulation for $|\bGam|=1$ which remains constant to numerical accuracy, according to its definition. 
}  
\end{figure}

\begin{figure}[h!]
\centering
\includegraphics[width=0.5\textwidth]{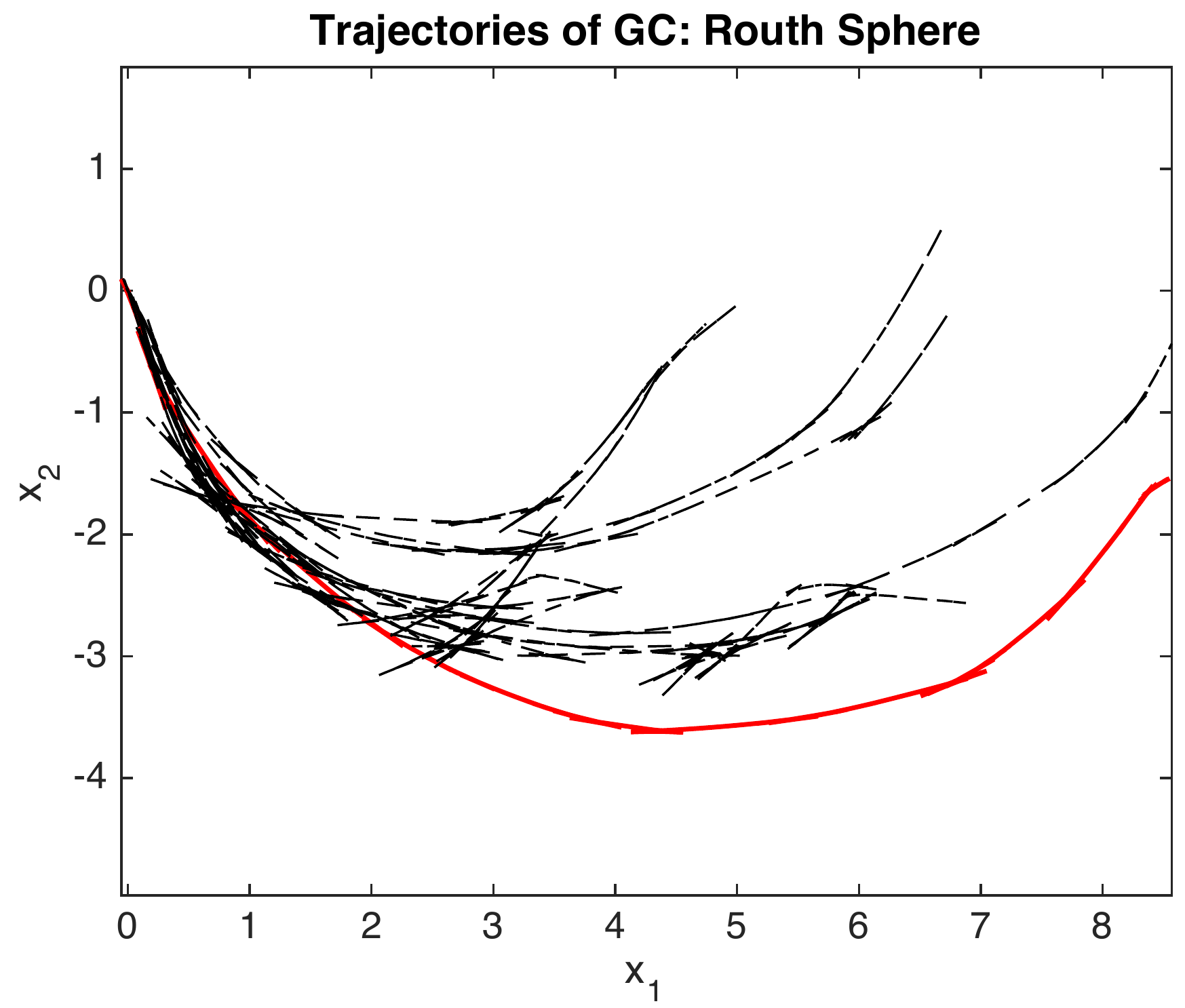} 
\caption{
\label{fig:Sim_Results_CG} 
Trajectories of the geometric center for the Routh Sphere corresponding to the computations presented in Figure~\ref{fig:Sim_Results}. Solid red line: noiseless trajectory; dashed lines: noise realizations. 
}  
\end{figure}

The non-conservation of energy, Jellet and Routh for stochastic rolling is verified in the numerical simulations shown in Figure \ref{fig:Sim_Results}.  This Figure also displays the preservation in the numerical simulations of the modulus of the unit vector $\bGam$. \revision{R1Q2n}{The evolution of the projection of the geometric center of the rolling ball is also shown on  Figure~\ref{fig:Sim_Results_CG}, computed from the formula \eqref{xgc_eq} derived earlier.  }
\revision{R1Q2l}{For simulations, we utilized the fully implicit Strong Stratonovich Euler-Heun  numerical method for computation of stochastic systems  \cite{GiSh2007}, implemented in \texttt{MATLAB}. We refer the reader to that publication and also \cite{FGBPu2015} for the details of numerical implementation of the method for the rolling sphere. }
\begin{remark} 
For most initial conditions and parameter values, we have observed numerically a nearly affine relationship between Jellet and Routh integrals, 
$R \simeq a J +b$, where $(a,b)$ depend on the parameter values.  This means, in principle, that there may be a constant of motion which we have not been able to identify yet. 
\end{remark} 

\revision{}{} 

\begin{figure}[h!]
\centering
\includegraphics[width=1\textwidth]{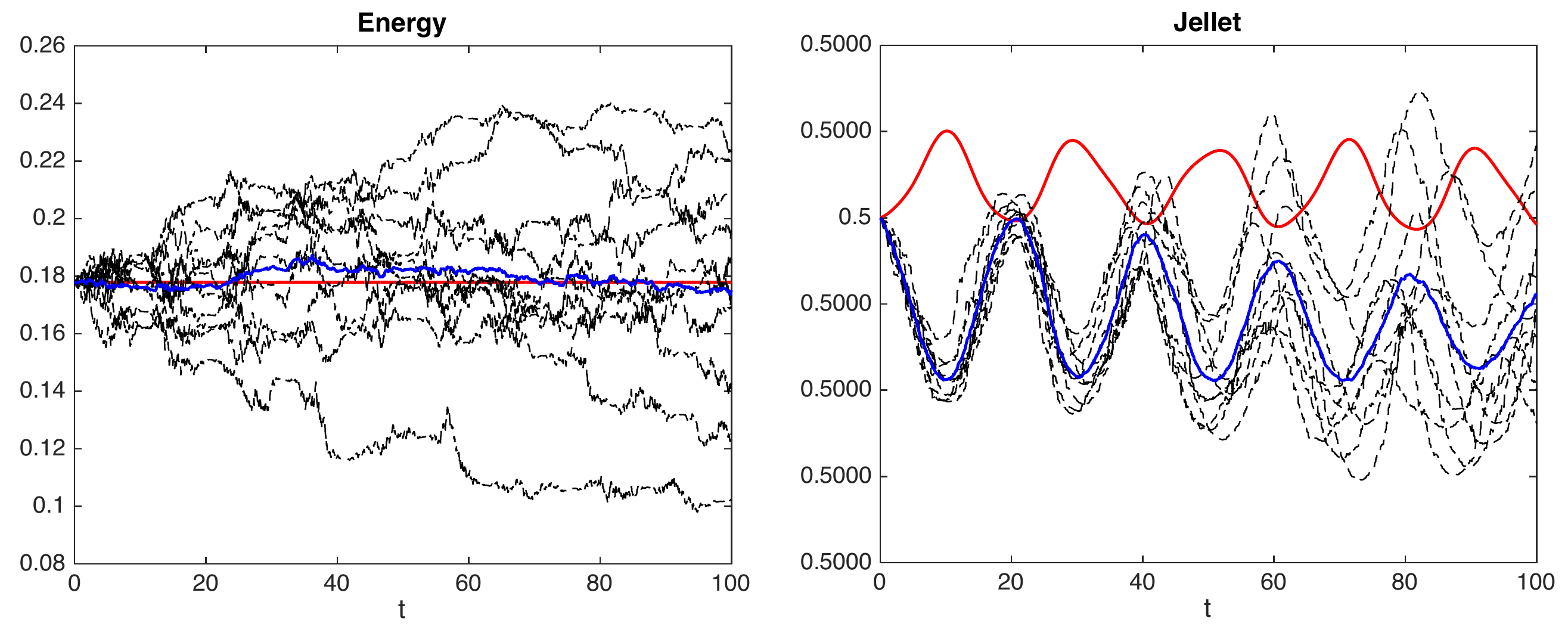} 
\caption{
\label{fig:Sim_Results_Chaplygin} 
Results of verification of simulation accuracy with $\bxi = 0.1 \bGam$, corresponding to vertical stochastic forcing from the substrate onto the ball, with the time step $\Delta t=0.025$.  Left panel: energy; right panel: $\mathbf{M} \cdot \bGam$, which is the analogue of the Jellet integral for Chaplygin's ball. Dashed black lines: different noise realizations. Blue solid line: mean values of different noise realizations. Red line: no noise (deterministic system).  As in Figure~\ref{fig:Sim_Results}, $|\bGam|$ is conserved to high precision ($10^{-6}$) and is not presented here. As expected from \eqref{Jellet_non_cons}, the analogue of Jellet integral $\mathbf{M} \cdot \bGam$ is conserved. }  
\end{figure}

\revision{R2Q11}{\subsection{Stochastic rolling of Chaplygin's ball}}
A simpler case that has been also well-studied in the literature is known as the \emph{Chaplygin ball}, which arises in the case when 
 the center of mass of the rolling ball coincides with its \revision{R2Q10}{geometric center} \emph{i.e.},  $\ell=0$, for arbitrary moments of inertia $(I_1,I_2,I_3)$. In this case, the right-hand side of equation \eqref{Noisy_Ball_explicit2}  vanishes  and one recovers the equations of motion for {\bfi Chaplygin's ball}:
 
\begin{align}
\begin{split}
\Big( {\sf d}  + \widetilde{\bOm} \times  \Big) \boldsymbol{M} 
&= 
  m r^2 \sum_i \bxi_i  \circ dW^i(t)  \times 
  \big( \bGam \times \left( \bOm \times \bGam \right) \big) 
\,,\\ 
\Big( {\sf d}  + \widetilde{\bOm} \times  \Big) \boldsymbol{\Ga}  &= 0
\,,\quad
\boldsymbol{M} 
=  \mathbb{I}\boldsymbol{\Om} 
+ mr^2\,  \boldsymbol{\Ga \times ( \Om \times \Ga ) }
\,.
\end{split}
\label{Chaplygin-sphere}
\end{align}
This expression can be also obtained by setting $\widetilde{\bOm}= \bOm \mbox{d}t + \sum_i \bxi_i \circ \mbox{d} W^i$, leading to the following equations, in agreement with \eqref{Noisy_Ball_explicit2}, 
\begin{align}
\begin{split}
& \Big( {\sf d}  +  \bOm \mbox{d} t \times  \Big) \boldsymbol{M} 
= 
 -\sum_i \bxi_i  \circ dW^i(t)  \times \mathbb{I} \bOm
\,,\\ 
& \Big( {\sf d}  + \widetilde{\bOm} \times  \Big) \boldsymbol{\Ga}  = 0
\,,\quad
\boldsymbol{M} 
=  \mathbb{I}\boldsymbol{\Om} 
+ mr^2\,  \boldsymbol{\Ga \times ( \Om \times \Ga ) }
\,.
\end{split}
\label{Chaplygin-sphere3}
\end{align}
Equations \eqref{Chaplygin-sphere} for Chaplygin's ball, by definition, preserve $|\boldsymbol{\Ga}|^2$. 
The deterministic equations preserve all four of the quantities $|{\bGam}|^2$, $\boldsymbol{M} \cdot{\bGam}$, $|\boldsymbol{M} |^2$ and energy with $\ell\to0$. In contrast, \eqref{Chaplygin-sphere3} preserves neither the magnitude of total momentum $|\bM|$, nor the analogue of the Jellet integral $\bM \cdot \bGam$. Indeed, $|\bM|^2$ evolves according to 
\begin{align} 
\begin{split}
\frac{1}{2} {\sf d}|\mathbf{M}|^2   &=\mathbf{M} \cdot  {\sf d} \mathbf{M} = m r^2 \sum_i \bxi \circ \mbox{d} W^i \cdot \left(\mathbb{I} \bOm \times 
\left( \bGam \times \left( \bOm \times \bGam \right)\right) \right)  
\\ 
&=m r^2 \sum_i \bxi \circ \mbox{d} W^i \cdot \left( \mathbf{M} \times 
\left( \bGam \times \left( \bOm \times \bGam \right)\right) \right)
\end{split}
\label{Momentum_non_cons} 
\end{align} 
and the rate of change for the analogue of  Jellet $\bM \cdot \bGam$ is computed to be 
\begin{equation} 
 {\sf d} \left( \mathbf{M} \cdot \bGam  \right)  =\mathbf{M} \cdot  {\sf d} \bGam + \bGam \cdot {\sf d} \mathbf{M} = 
m r \sum_i \bxi_i \circ \mbox{d} W^i \cdot \left(  \bOm \times \bGam \right) .
\label{Jellet_non_cons} 
\end{equation} 
Since $\ell=0$, there is no analogue of Routh integral, as the choice of $\bchi$ is arbitrary. 
For the deterministic solution behaviour of Chaplygin's ball, see, e.g., \cite{Ki2001}. We see from \eqref{Jellet_non_cons} that if $\bxi$ is a time-independent vector in the body frame, the Jellet integral will not be conserved. However, as a test case for verifying our simulations, we may formally put $\bxi=0.1 \bGam$. Technically, this choice is inconsistent as $\bxi$ cannot depend on variables in the body frame. Nonetheless, this simulation is useful in illustrating the accuracy of our numerical schemes  and, thereby, verifying the correctness of our analysis.  We present the numerical solutions of Chaplygin's ball on Figure~\ref{fig:Sim_Results_Chaplygin}. As one can observe from \eqref{Jellet_non_cons}, in this case, the analogue of Jellet integral should be conserved, which is indeed illustrated on the right panel of the Figure. The left panel of the figure shows the evolution of energy, which is not conserved, as expected from the analysis. If we take $\bxi$ to be an arbitrary vector in the body frame, which is either constant or has a prescribed dependence on time, but not on the dynamical variables, then neither the energy nor the Jellet integral is conserved. 

\section{Summary}
\label{sec:summary} 

This paper has shown that stochasticity representing uncertainty in the angular velocity at which a nonholonomic rolling constraint is applied can have dramatic effects on the ensuing dynamics. In particular, this sort of stochasticity destroys the corresponding deterministic conservation laws and thereby liberates the solution to produce large deviations in wandering paths. Thus, nonholonomic constraints can amplify the effects of this sort of noise and create large uncertainty. The Hamilton-Pontryagin approach taken here has been shown to possess an equivalent Lagrange-d'Alembert counterpart. Consequently, all of these results are also available from an alternative viewpoint in the more traditional Lagrange-d'Alembert approach. Moreover, all of the properties associated with geometric mechanics, \revision{R2Q15}{such as reduction by symmetry,} are retained in both approaches. 

We expect it will be interesting in future work to see how the underlying geometric framework for deterministic nonholonomic systems will be used to characterize the probabilistic aspects of their solution behaviour when stochasticity is introduced into the nonholonomic constraints. 
For example, the relationship between stochastic variational methods and other approaches to introducing stochasticity in mechanical systems offers opportunities for further development. To be more concrete, let us come back to the physical question of experimental observation of a  ball rolling on a table. Suppose that the ball is traced using several features on its surface, \emph{e.g.} bright dots, using a camera. To track the ball's dynamics, one can either determine the position of the dots and infer its orientation, compute the velocity of the dots in space and infer the angular velocity of the ball, combine these methods with information provided by the linear velocity of the ball and use the rolling condition, or  perhaps even employ an approach combining all of the above methods. Each of these techniques will lead to a different stochasticity in equations, and some of these measurement methods may not be of Stochastic Transport (ST) type considered here. One may also be interested in combining  stochastic extensions of nonholonomic rolling conditions as in \cite{FGBPu2015} with the ST noise.  Making this combination represents another interesting and challenging problem which should be treatable by our methods. 
\revision{R1Q2o}{Yet another conceivable endeavour would be to use the present formalism in developing control methods for nonholonomic systems with errors in location or velocity, e.g., moving on rough terrain and/or experiencing random slippage. This endeavour might be interesting for the development of practical rolling robots.}  We believe that further considerations and combinations of different types of stochastic dynamics in nonholonomic systems will be interesting and important. We  will consider these endeavours in our future work. 

\subsection*{Acknowledgements}
We are enormously grateful to our friends and colleagues whose remarks and responses have encouraged us in this work. We thank  A. A. Bloch, F. Gay-Balmaz, M. Leok, T. S. Ratiu,  D. V. Zenkov and many others who have offered their valuable suggestions in the course of this work.  

\subsection*{Funding Statement} The work of DDH was partially supported by by the European Research Council Advanced Grant 267382
FCCA and EPSRC Standard Grant EP/N023781/1. The work of VP was partially supported by the NSERC Discovery grant and University of Alberta's Centennial Professorship.

\subsection*{Author contribution } 
Both authors contributed equally to the derivation of theoretical models, their analysis and interpretation of numerical results.

\subsection*{Competing interests} The authors have no competing interests.

\subsection*{Ethics statement} This work did not involve any collection of human data. 

\subsection*{Data accessibility } This work does not have any experimental data.

{\footnotesize 
\bibliographystyle{unsrt}
\bibliography{Rolling-Ball-References,bibliography}
}

\appendix
\section{Derivation of evolution for Jellet and Routh integrals in the stochastic case} 
\label{app:Jellet_Routh} 
\subsection*{Jellet integral} 
The evolution equations for Jellet integral $J = \mathbf{M} \cdot \bs$ are obtained as follows. 
\begin{equation} 
\label{dJ_expression} 
\begin{aligned} 
\mbox{d} J & = \mathbf{M} \cdot \mbox{d} \bs + \bs \cdot \mbox{d}  \mathbf{M} 
\\
&= \mathbf{M} \cdot \big( - \bOm \mbox{d} t \times r \bGam \big) 
-
 \mathbf{M} \cdot \left( \sum_i \bxi_i \circ \mbox{d} W^i \times r \bGam \right) 
\\ 
& \quad + 
\bs \cdot \big( - \bOm \mbox{d} t \times \mathbf{M}  \big) 
+ 
m g \ell \bs \cdot \left( \bGam \times \bchi \right)  \mbox{d} t 
\\
& \quad 
+ \bs \cdot \big( m \bY \times \left( \bOm \times r \bGam \right) \big)\mbox{d} t
 -
  \bs \cdot \big(\sum_i \bxi_i \circ \mbox{d} W^i \times \left( \mathbf{M}-r \bGam \times m \bY \right)  \big) 
 \\ 
 & = 
 - \big( \bs - r \bGam \big) \big( \bOm \times \mathbf{M} \big) \mbox{d} t 
 + 
 \big( \bs \times m \bY \big) \cdot \big(\bOm \times r \bGam \big) \mbox{d} t
 \\
 & \quad 
 + 
 \sum_i \bxi_i \circ \mbox{d} W^i \big( \ell \bchi \times \mathbf{M} + \left( \bGam \times m \bY \right) \times \bs \big) 
 \\
 & = 
\quad   \big( \ell \bchi  + r \bGam \big) \big( \bOm \times \left( \bs \times \m \bY \right) \big) \mbox{d} t + 
\ell \bchi \cdot \big( \bOm \times \mathbb{I} \bOm \big)  \mbox{d} t 
\\& \quad 
+ 
\sum_i \bxi_i \circ \mbox{d} W^i \cdot\left( 
\ell \bchi \times 
\mathbb{I} \bOm + \ell \bchi \times \big( \ell \bchi \times m \bY \big) + \big(r \bGam \times m \bY \big) \times r \bGam 
\big) 
\right) 
\\
&=\sum_i \bxi_i \circ \mbox{d} W^i \cdot \left( 
\ell \bchi \times 
\mathbb{I} \bOm + \ell \bchi \times \big( \ell \bchi \times m \bY \big) + \big(r \bGam \times m \bY \big) \times r \bGam 
\big) 
\right) 
 \,,
\end{aligned} 
\end{equation} 
Here we have used the identity $\bs \cdot (\bchi \times \bGam ) =0$, and also noticed that $\bchi \cdot (\bOm \times \mathbb{I} \bOm )=0$ for $I_1=I_2$. 

\subsection*{Routh integral}
We remind the reader that the Routh integral can be written in our notation as $R=\Omega_3 \sqrt{I_1 I_3 + m \mathbb{I} \bs \cdot \bs}$. In order to find the evolution equation for  this quantity in the stochastic case, we start with finding the $\bchi$-projection of equation \eqref{Noisy_Ball_explicit1} by computing the following:
\begin{equation} 
\begin{aligned} 
\bchi \cdot & \left( \mathbb{I} \Omega + \bs \times m \bY \right)  = I_3 \Omega_3 + m |\bs|^2 \Omega_3 - s_3 \bOm \cdot \bs 
\\
& = I_3 \Omega_3 + m |\bs|^2 \Omega_3 -\frac{1}{I_1}  s_3 \left( J-\big(I_3-I_1 \big) \Omega_3 s_3 \right)  = 
\frac{1}{I_1} \left( I_1 I_3 + m \mathbb{I} \bs \cdot \bs \right) \Omega_3 + \frac{1}{I_1} J s_3 
\\ 
{\sf d} s_3 & = \bchi \cdot {\sf d} \bs = \bchi \cdot \left( - \widetilde{\bOm} \times r \bGam \right)  
\\ 
& =  
 \bchi \cdot \left( - \bOm {d} t  \times \bs \right) + \bchi \cdot \left( \sum_i \bxi_i \circ \mbox{d} W^i \times \bs \right)  
 = - Y_3\,dt - \sum_i \bxi_i  \circ \mbox{d} W^i  \cdot \left( \bs \times \bchi \right) 
 \\
 \mathbb{I} \bs &= {\rm diag}(I_1,I_1,I_3) \bs = \left( I_1 s_1, I_1 s_2, I_3 s_3 \right) = I_1 \bs + \left( I_3-I_1 \right) s_3 \bchi 
 \\ 
 \frac{1}{2}  {\sf d} \mathbb{I} &\bs \cdot \bs = 
 \left( I_1 \bs + \big( I_3 - I_1 \big) s_3 \bchi  \right) \cdot \left( - \widetilde{\bOm} \times \big( \bs - \ell \bchi \big) \right) \\ 
 & = I_1 \bs \cdot( \widetilde{\bOm} \times l \bchi ) + (I_3-I_1) s_3 \bchi \cdot ( - \widetilde{\bOm} \times \bs ) 
 = 
 \left( I_1 l + (I_3-I_1) s_3 \right) \bchi \cdot (-  \widetilde{\bOm} \times \bs ) 
 \\ & =  - \left( I_1 l + (I_3-I_1) s_3 \right) \left( Y_3 {d} t + \sum_i \bxi_i\circ \mbox{d} W^i \cdot ( \bs \times \bchi ) \right) 
 \\ 
 J &= \mathbf{M}\cdot \bs = \mathbb{I} \bOm \cdot \bs = I_1 \bOm \cdot \bs  + \big( I_3-I_1 \big) \Omega_3 s_3 
 \\ 
 \bchi \cdot & \left( \bOm \times \left( \bs \times m \bY \right)  \right) = -m \bchi \cdot \bY (\bOm \cdot \bs ) = 
 -m Y_3 \left(\frac{J}{I_1} - \frac{I_3-I_1}{I_1} \Omega_3 s_3  \right) 
 \\ 
  \bchi \cdot &  \left( m \bY \times \left( \bOm \times l \bchi \right) \right) = 
  m \bchi \cdot \bOm ( \bY \cdot l \bchi ) = m l \Omega_3 Y_3 
\end{aligned} 
\label{useful_identity_Routh}
\end{equation} 
Here, we used the expression for Jellet integral $J = \mathbb{I} \bs \cdot \bs$ derived in \eqref{dJ_expression}, and introduced a short-hand notation, $Y_3 = \bchi \cdot \bY$. We multiply \eqref{Noisy_Ball_explicit2} by $\bchi$ and compute using 
\eqref{useful_identity_Routh} as follows 
\begin{equation} 
\label{Routh_deriv} 
\begin{aligned} 
{\sf d}& \bchi \cdot \left( \mathbb{I} \Omega + \bs \times m \bY \right) = 
\frac{1}{I_1}{\sf d} \Big[ \left( I_1 I_3 + m \mathbb{I} \bs \cdot \bs \right) \Omega_3 \Big]+ \frac{s_3}{I_1} {\sf d} J + \frac{J}{I_1} {\sf d} s_3  
\\ 
& = \frac{1}{I_1}{\sf d} \big[ \left( I_1 I_3 + m \mathbb{I} \bs \cdot \bs \right) \Omega_3 \big]+ \frac{s_3}{I_1} {\sf d} J 
-\frac{J}{I_3} \left(  Y_3\,{\sf d}t - \bxi_i \circ {\sf d} W^i \cdot \left( \bs \times \bchi \right)  \right) 
\\
& = 
- \frac{J}{I_1} \left( m Y_3   - \frac{I_3-I_1}{I_1} \Omega_3 s_3 \right) {\sf d} t+ m \ell \Omega_3 Y_3
\\ & \qquad  + 
\sum_i \bxi_i \circ {\sf d} W^i \cdot \big[ \bchi \times \left( \mathbb{I} \bOm + \ell \bchi \times m \bY \right) \big] 
\end{aligned} 
\end{equation} 
After using additional identities from \eqref{useful_identity_Routh}, we get 
\begin{equation} 
\label{Routh_deriv2} 
\begin{aligned} 
\frac{1}{I_1}& {\sf d} \big[ \left( I_1 I_3 + m \mathbb{I} \bs \cdot \bs \right) \Omega_3 \big]+\frac{1}{2} m \Omega_3 {\sf d} \big( \mathbb{I} \bs \cdot \bs \big) 
\\
&=- \frac{s_3}{I_1} {\sf d} J 
+ 
\sum_i\bxi_i \circ {\sf d} W^i \cdot \left[ \bchi \times \left\{ \left( \mathbb{I} \bOm + \ell \bchi \times m \bY \right) - 
m \bs \frac{I_1 \ell + (I_3-I_1) s_3}{I_1}  \right\} \right] 
\end{aligned} 
\end{equation} 
Finally, multiplying \eqref{Routh_deriv2} by $\Omega_3$, we obtain 
\begin{equation} 
\label{Routh_cons} 
\frac{1}{2} {\sf d} R^2 = \frac{1}{2} {\sf d} \left[ \Omega_3^2 \left( I_1 I_3 + m \mathbb{I} \bs \cdot \bs \right) \right] = 
- \frac{s_3 \Omega_3}{I_1} {\sf d} J + 
\sum_i \bxi_i \circ {\sf d} W^i \cdot \mathbf{F}(\bOm, \bs) 
\end{equation} 
where we have defined for brevity of notation, 
\begin{equation} 
\label{Fdef} 
\begin{aligned} 
 \mathbf{F}(\bOm, \bs) & = \Omega_3  \bchi \times  \Big[
 \left( \mathbb{I} \bOm + \ell \bchi \times m \bY \right) - 
m \bs \frac{I_1 \ell + (I_3-I_1) s_3}{I_1} 
\Big] 
\end{aligned} 
\end{equation} 
and used the formula for the stochastic evolution of the Jellet integral ${\sf d} J$ given by \eqref{dJ_expression}.

\end{document}